\documentclass[aps,prl,twocolumn,superscriptaddress,longbibliography,nobalancelastpage,floatfix]{revtex4-2}
\pdfoutput=1 
\usepackage{graphicx}
\usepackage{amsmath}
\usepackage{amssymb}
\usepackage[explicit]{titlesec} 
\usepackage{bm}
\usepackage{physics}
\usepackage[colorlinks = true,
            allcolors = {Blue}]{hyperref}
\usepackage[capitalize]{cleveref}
\usepackage[caption=false]{subfig}
\usepackage[dvipsnames]{xcolor}

\begin{document}

\title{Hidden nonlinear optical susceptibilities in linear polaritonic spectra}

\author{Arghadip Koner}
\author{Joel Yuen-Zhou}%
 \email{joelyuen@ucsd.edu}
\affiliation{%
 Department of Chemistry and Biochemistry, University of California San Diego, La Jolla, California 92093
}%
\date{\today}

\begin{abstract}

Linear spectra of molecular polaritons formed by $N$ molecules coupled to a microcavity photon mode are usually well described by classical linear optics, raising the question of where the expected nonlinear effects in these strongly coupled systems are. In this work, we derive a general expression for the polaritonic linear spectra that reveal previously overlooked finite-size quantum corrections due to vacuum-mediated molecular Raman processes. Using a $1/N$ expansion, we demonstrate that these nonlinearities are suppressed in typical low-Q cavities due to an emergent timescale separation in polariton dynamics yet manifest in high-Q single-mode cavities where the photon loss is comparable to the single-molecule light-matter coupling strength.

\end{abstract}

\maketitle

\noindent Polaritons have garnered significant attention in the last decade due to their promise to manipulate matter degrees of freedom (DOF) via coupling to a confined electromagnetic mode. They have offered a plethora of promising applications like enhanced energy transport~\cite{coles2014polariton,Reitz2018energy,koner2023path}, modification of chemical reactions ~\cite{hutchison2012modifying,herrera2020molecular}, and room temperature lasing~\cite{Freire2024room,Bennenhei2023polarized}. For molecular systems, strong coupling (SC) is typically realized through the interaction of an ensemble of \(N \approx 10^6 - 10^{12}\) molecules with a microcavity mode~\cite{daskalakis2017polariton}. The multiple exchanges of excitation between the electromagnetic mode and the matter optical polarization (see Fig~\ref{fig:schematic}a) are expected to induce nonlinear optical transitions in the molecules. However, recent studies suggest that several polaritonic phenomena can be replicated using appropriately shaped lasers, causing polaritons to sometimes act just as optical filters~\cite{schwennicke2024filtering}. This effect is evident in the linear response of polaritons, which is well described using classical linear optics methods, such as transfer matrices, with the molecular optical constants as the only material input~\cite{zhu1990vacuum, schubert1996polarization, yariv2007photonics, nemet2020transfer, cwik2016excitonic, gunasekaran2023continuum, yuen2023linear}. This classical linear optics description raises a fundamental question: what happened to the nonlinearities anticipated in the SC regime? The answer to this question is also key to uncovering genuine cavity-induced effects beyond classical optics.\\
In this letter, we solve this apparent contradiction by obtaining a general expression for the linear response of molecular polaritons, which evidences that the classical optical description is only correct in the thermodynamic limit ($N\rightarrow\infty$) and, in fact, there is a previously overlooked hierarchy of (finite-size) $1/N$ corrections containing signatures of nonlinear molecular processes like Raman scattering mediated by quantum vacuum fluctuations of the cavity. These corrections are small due to an emergent separation of timescales in polariton dynamics and are usually excluded from the spectra due to limited resolution in the low-Q cavities typically used in experiments. They manifest in high-Q single-mode cavities where the cavity decay rate $\kappa$ becomes comparable to the single molecule coupling $\lambda$. \\
\begin{figure}
\begin{centering}
\includegraphics[width=0.5\linewidth]{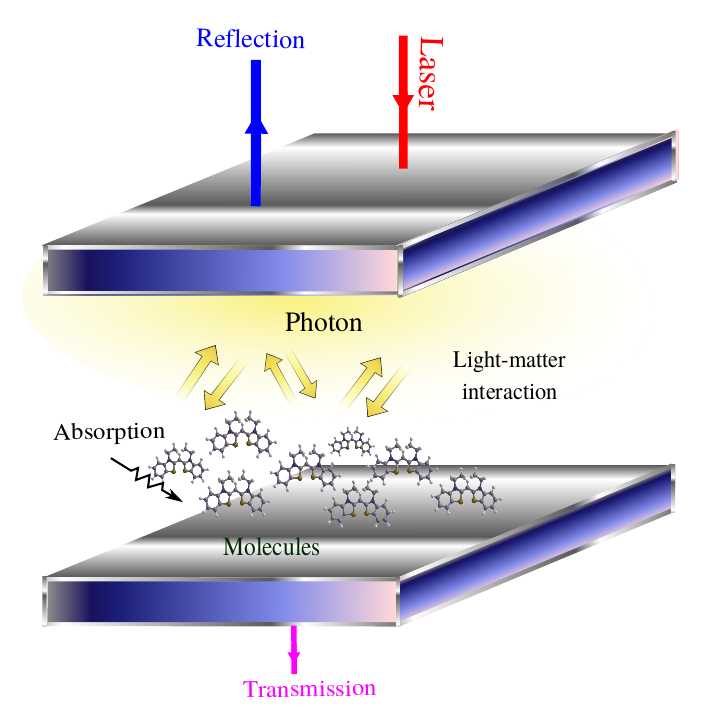}
\par\end{centering}
\caption{Strong coupling between a cavity photon and molecular optical polarization leads to multiple exchanges of excitation between light and matter. We explore the conditions under which these exchanges lead to nonlinear optical processes in the molecules, even at linear order in the incoming photon field. \label{fig:schematic}\label{fig:schematic}} \end{figure}

\noindent \textbf{Model.}\textemdash  We consider a photonic cavity with a single-mode of frequency $\omega_{\text{ph}}$ coupled to $N$ non-interacting molecules. Starting from a permutationally invariant initial state at zero temperature, the setup is described by a bosonic Hamiltonian at all times [Supplemental material: Sec. \textcolor{blue}{S1}]~\cite{perez2024bosons,Richter2016efficient,Shammah2018open,Keeling2022efficient,Silva2022permutational,Sukharnikov2023second},
\begin{eqnarray}\label{bosonic_Hamiltonian}
   \nonumber H&=& \bigg[\hbar\omega_{\text{ph}}a^\dagger a + \hbar \sum_{j} \omega_{g,j} b_j^\dagger b_j + \hbar \sum_{j} \omega_{e,j} B_j^\dagger B_j\bigg]+ \\  
    & &  \bigg[-\hbar \lambda \sum_{jj'} \langle \varphi^{(e)}_{j'}| \varphi^{(g)}_{j} \rangle ab_j B_{j'}^\dagger - \text{h.c.}\bigg],
\end{eqnarray}
with operators $a$ that annihilates a cavity excitation (photon), and $b_j(B_{j'})$ that annihilate a molecule in vibronic state $|g, \varphi^{(g)}_{j}\rangle$($|e, \varphi^{(e)}_{j'}\rangle$); $j(j')$ label the vibrational level in the ground (excited) electronic state. To compute linear response, we project the Hamiltonian to the first excitation manifold, where it admits a block tridiagonal representation~\cite{perez2024cute}:\small
\begin{equation}\label{CUT-E_matrix}
H_1=\begin{pmatrix}
\bm{H}_{\text{ph},0} & \bm{V}_{0} & 0 & \dots & 0 & 0 & 0 \\
\bm{V}_{0}^{\dagger} & \bm{H}_{e,0} & \bm{v}_{0} & \dots & 0 & 0 &  \\
0 & \bm{v}_{0}^{\dagger} & \bm{H}_{\text{ph},1} & \bm{V}_{1} & 0 & 0 & 0 \\
\vdots & \vdots & \vdots & \ddots & \vdots & \vdots & \vdots \\
0 & 0 & 0 & \dots & \bm{V}_{N-1}^{\dagger} & \bm{H}_{e,N-1} & \bm{v}_{N-1} \\
\vdots & \vdots & \vdots & \vdots & 0 & \bm{v}_{N-1}^{\dagger} & \bm{H}_{\text{ph},N}
\end{pmatrix}.
\end{equation}
\normalsize
Here $\bm{H}_{\text{ph},n}$ denotes the effective cavity single-photon Hamiltonian for the subspace where $n$ molecules have phonons in the ground state (e.g. $\bm{H}_{\text{ph},0}=(\omega_{\text{ph}}-i\kappa/2)|1_{\text{ph},0}\rangle\langle 1_{\text{ph},0}|$ for a cavity excitation with all molecules in $|g,\varphi_0^{(g)}\rangle$), and  $\bm{H}_{e,n}$ corresponds to the photonless subspace with one molecule excited and $n$ other molecules having phonons in the ground state~\cite{Philpott1971theory,
Herrera2017Dark}. The off-diagonal elements reveal the crucial timescale separation: couplings $\bm{V}_{n}$ drive the collective $\mathcal{O}(\lambda\sqrt{N})$ `Rayleigh' transitions that conserve the number of molecules with ground state phonons, and $\bm{v}_{n}$ mediate the single-molecule $\mathcal{O}(\lambda)$ `Raman' transitions that create/destroy the phonons in one of the molecules~\cite{perez2023simulating}.\\

 The linear spectroscopic observables of cavity-coupled systems (absorption, reflection, transmission) can be entirely expressed in terms of the photon Green's function $D_N^{R}(\omega) =\langle 1_{\text{ph,0}} |\bm{G}(\omega)|1_{\text{ph,0}}\rangle$ [Supplemental material: Sec. \textcolor{blue}{S2}], where $\bm{G}(\omega)=\frac{1}{\omega-H_1+i0^+}$. In particular, the absorption is given by ~\cite{yuen2023linear}, 
\begin{align}
A(\omega) & =-(\kappa/2)[\kappa|D_N^{R}(\omega)|^{2}+2\Im D_N^{R}(\omega)],\label{eq:A_w}
\end{align}
To compute $D_N^{R}(\omega)$, we exploit the block tridiagonal structure of $H_1$ and use standard techniques for matrix Green's functions of systems with nearest neighbor couplings [see Supplemental material: Sec. \textcolor{blue}{S3} ]~\cite{Velev2004gf,pastawskii1983matrix,Hangii1978continued}. Rewriting the Hamiltonian $H_1=\begin{pmatrix}\bm{H}_{\text{ph},0} & \bm{V}_{0}\\
\bm{V}_{0}^{\dagger} & \tilde{\bm{H}}_{e,0}
\end{pmatrix}$
yields
\begin{eqnarray}\label{DRw1}
    D_N^{R}(\omega)=\frac{1}{\omega-\omega_{\text{ph}}+i\kappa/2-\Sigma_{e,0}(\omega)},
\end{eqnarray}
where $\Sigma_{e,0}(\omega)=\bm{V}_{0}(\omega-\tilde{\bm{H}}_{e,0}+i\gamma/2)^{-1}\bm{V}_{0}^{\dagger}$ is the photon self-energy due to its coupling to the $\tilde{\bm{H}}_{e,0}$ block; here $\gamma\rightarrow 0$ is introduced to ensure causality of the Green's function~\cite{MukamelBook}. We next obtain an expression for $\Sigma_{e,0}(\omega)$ in terms of the bare Green's functions and the off-diagonal light-matter couplings in Eq.~\ref{CUT-E_matrix}. By iterating the technique used to obtain Eq.~\ref{DRw1} [Supplemental material: Sec. \textcolor{blue}{S4}], we arrive at a recursive relation where the self-energy at the $k^{\text{th}}$ step depends on the $(k+1)^{\text{th}}$ step~\cite{Velev2004gf,pastawskii1983matrix}:\small{
\begin{subequations}\label{Eq:recursion}
\begin{eqnarray}
&\boldsymbol{\Sigma}_{e,k}=&\bm{V}_{k}\big(\omega-\bm{H}_{e,k}+i\gamma/2-\boldsymbol{\Sigma}_{\text{ph},k+1}\big)^{-1}\bm{V}_{k}^{\dagger},\\
&\boldsymbol{\Sigma}_{\text{ph},k+1}=&\bm{v}_{k}\big(\omega-\bm{H}_{\text{ph},k+1}+i\kappa/2-\boldsymbol{\Sigma}_{e,k+1}\big)^{-1}\bm{v}_{k}^{\dagger},
\end{eqnarray}
\end{subequations}}\normalsize
\\
for $0 \leq k < N$. For finite $N$, the series truncates at $\boldsymbol{\Sigma}_{\text{ph},N}=\bm{v}_{N-1}\big(\omega-\bm{H}_{\text{ph},N}+i\kappa/2\big)^{-1}\bm{v}_{N-1}^{\dagger}$. Thus Eq.~\ref{DRw1} becomes,
\begin{widetext}
\begin{equation}\label{Eq:contd_frac_full}
D^R_N(\omega)=\frac{1}{\omega-\omega_{\text{ph}}+i\kappa/2-\bm{V}_{0}\frac{1}{\omega-\bm{H}_{e,0}+i\gamma/2-\bm{v}_{0}^{\dagger}\frac{1}{\omega-\bm{H}_{\text{ph,1}}+i\kappa/2-\bm{V}_{1}^{\dagger}\frac{1}{\frac{\ddots}{\omega-\bm{H}_{e,N-1}+i\gamma/2-\bm{v}_{N}^{\dagger}\frac{1}{\omega-\bm{H}_{\text{ph,}N}+i\kappa/2}\bm{v}_{N}}}\bm{V}_{1}}\bm{v}_{0}}\bm{V}_{0}^{\dagger}}.
\end{equation}
\end{widetext}

\noindent \textbf{$1/N$ expansion.}\textemdash  Eq.~\ref{Eq:contd_frac_full} admits a $1/N$ expansion due to the aforementioned timescale separation in the polariton dynamics, $D_N^R(\omega)=\sum_{k=0}^\infty d_{N,k}(\omega)$ where $d_{N,k}(\omega)\propto N^{-k}$.
Here, we present closed-form expressions for $d_{N,0}$ and $d_{N,1}$. For $k\ge 2$, certain non-commuting cascade processes prevent similar analytical manipulations (Supplemental material: Sec. \textcolor{blue}{S6}), but we can still find expressions for $d_{N,k}$ correct up to $\mathcal{O}(N^{-k})$.
 \begin{figure*}
\begin{centering}
\includegraphics[width=0.9\linewidth]{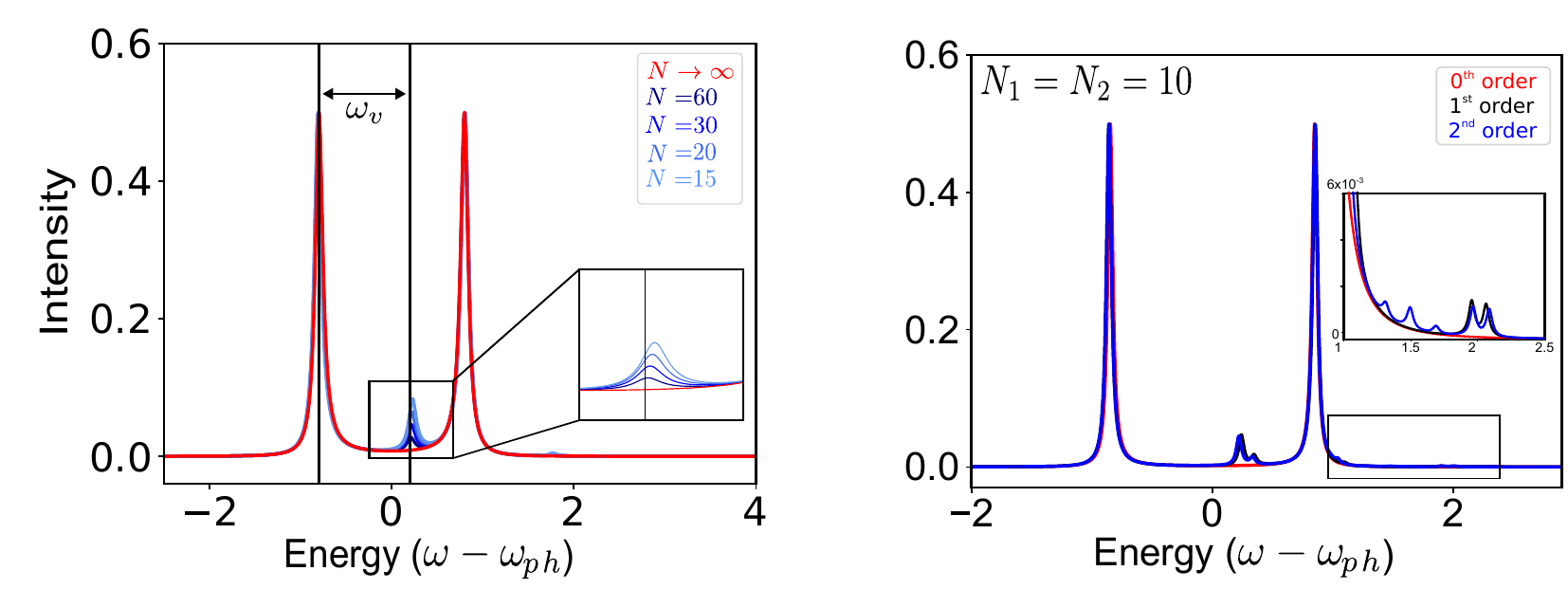}
\par\end{centering}
\caption{ (a) Linear absorption spectrum of an ensemble of 3-level systems (two ground-state vibrational levels and one excited level) at zero temperature, with fixed collective coupling $\lambda\sqrt{N}$ while varying $N$. The red curve is the thermodynamic limit result $(N\rightarrow\infty)$ and is consistent with classical linear optics; it features resonances at the zeroth-order polaritons. First-order polariton sidebands show up at a vibrational frequency $\omega_v$ to the blue of each zeroth-order polariton and indicate Raman processes mediated by vacuum (hence, going beyond classical optics). The height of these sidebands increases as $N$ decreases due to increasing single molecule coupling $\lambda$. Parameters: $\omega_v = 1$, $\kappa = \gamma = 0.1\omega_v$, $\lambda\sqrt{N} = 0.8\omega_v$, $\omega_{\text{ph}} = \omega_{\text{e,0}}$. (b) Linear absorption of an ensemble of two types of 3-level molecules with different vibrational gaps but the same electronic gap, including up to $\mathcal{O}(N^{-2})$ corrections. The red and black curves show zeroth- and first-order polaritons as in (a). The blue curve shows $\mathcal{O}(N^{-2})$ Raman corrections to the spectrum. Parameters: $\omega_{v,A} = 1$, $\omega_{v,B} = 1.2$, $\kappa = \gamma = 0.05\omega_v$, $\lambda\sqrt{N} = 0.6\omega_v$, $\omega_{\text{ph}} = \omega_{\text{e,0}}$ for both molecules.\label{fig:results}}
\end{figure*}
From Eq.~\ref{Eq:contd_frac_full}, it is easy to see that $d_{N,0}$ must have no $\bm{v}_n^\dagger$ or $\bm{v}_n$ terms, so, 
\begin{align}\label{zeroth_DR}
d_{N,0}(\omega)
&=\frac{1}{\omega-\omega_{\text{ph}}+i\kappa/2-\bm{V}_{0}\bm{G}_{e,0}(\omega)\bm{V}_{0}^{\dagger}},
\end{align}
where $\bm{G}_{e,0}$ is the bare Green's function of $\bm{H}_{e,0}$ and hence $-\bm{V}_{0}\bm{G}_{e,0}(\omega)\bm{V}_{0}^{\dagger} \equiv \frac{\omega_{\text{ph}}}{2}\chi^{(1)}(\omega)=-\text{lim}_{\gamma\to0^{+}}N \sum_{j,j'}|\lambda|^2\frac{|\langle g,\varphi_j^{(g)}|\mu|e,\varphi_j'^{(e)}|^{2}}{\omega-(\omega_{e,j'}-\omega{g,j})+i\frac{\gamma}{2}}$, where $\chi^{(1)}(\omega)$ is the standard linear susceptibility of the molecular ensemble for a sample of density $N/\mathcal{V}_{\text{mol}}$ with $\mathcal{V}_{\text{mol}}$ equal to the cavity mode volume $\mathcal{V}_{\text{ph}}$, and $\mu$ as the molecular transition dipole operator. The poles of $d_{N,0}(\omega)$ will hereafter be referred to as the zeroth-order polariton frequencies. In the thermodynamic limit ($N\to \infty$ or $\lambda\to0$ while $\lambda\sqrt{N}$ is constant), $D^R_N(\omega)\approx d_{N,0}(\omega)$; so the linear response of polaritons is compactly expressed in terms of the material linear susceptibility, a result that is consistent with classical optics~\cite{cwik2016excitonic, yuen2023linear, schwennicke2024filtering}.\\

We can proceed analogously with the $1/N$ correction [Supplemental material: Sec. \textcolor{blue}{S6}],
\begin{widetext}
   \begin{eqnarray}\label{DR_1N}
d_{N,1}(\omega)  
   &=d_{N,0}(\omega) \cdot\bm{V}_{0}\bm{G}_{e,0}\bm{v}_{0}\big(\omega-   \bm{H}_{\text{ph},1}+i\kappa/2 -\bm{V}_{1}\bm{G}_{e,1}\bm{V}_{1}^{\dagger}\big)^{-1}\bm{v}_{0}^{\dagger}\bm{G}_{e,0}\bm{V}_{0}^{\dagger}\cdot d_{N,0}(\omega). 
\end{eqnarray} 
\end{widetext}
In the expression, we refer to the poles of $\big(\omega-\bm{H}_{\text{ph},1}+i\kappa/2 -\bm{V}_{1}\bm{G}_{e,1}\bm{V}_{1}^{\dagger}\big)^{-1}$ as the first-order polariton frequencies, which are shifted from the zeroth-order polaritons by a Raman (vibrational) transition. Eq.~\ref{DR_1N} can be interpreted from right to left as a sequential Stokes-anti-Stokes Raman scattering process: (a) The photon entering the cavity through the zeroth-order polariton $d_{N,0}(\omega)$ carries out a Stokes Raman transition in one of the $N$ molecules via $\bm{v}_0^\dagger\bm{G}_{e,0}\bm{V}_0^\dagger$, (b) the leftover Stokes photon strongly couples to the $(N-1)$ remaining molecules forming first-order polaritons, where the Raman excited molecule serves as a spectator; (c) the first-order polariton now induces the reverse (anti-Stokes) Raman transition resetting all the molecules back to their electronic and vibrational ground state via $\bm{V}_0\bm{G}_{e,0}\bm{v}_0$, while releasing the photon to the zeroth-order polariton $d_{N,0}(\omega)$. Since the creation and the annihilation of the Raman phonon are cavity vacuum-mediated, these transitions are individually penalized by $1/\sqrt{N}$ and go beyond the classical optics formalism. To understand the spectroscopic fingerprints, we present an illustrative example.\\
\noindent Let us consider the model in Eq.~\ref{bosonic_Hamiltonian}, with each molecule having two vibrational levels in the ground state ($|g,\varphi^{(g)}_{0}\rangle$, $|g,\varphi^{(g)}_{1}\rangle$) and one excited state ($|e,\varphi^{(e)}_{0}\rangle$); this simplified model is already rich enough to illustrate the physics of interest. 
The vibrational gap is $\omega_v=\omega_{\varphi^{(g)}_{1}}- \omega_{\varphi^{(g)}_{0}}$ and
the cavity is resonant with the $|g,\varphi^{(g)}_{0}\rangle \rightarrow |e,\varphi^{(e)}_{0}\rangle$ transition with frequency $\omega_{e,0}$. Fig.~\ref{fig:results}a shows the $A(\omega)$ in the thermodynamic limit and the $1/N$ corrections computed using Eq.~\ref{DR_1N}. As $N$ decreases, the $1/N$ correction reveals two sidebands separated from the zeroth-order polaritons by $\omega_v$. These first-order polaritons have a Rabi splitting of $2\lambda\sqrt{N-1}$, and they have featured in previously reported calculations of polaritonic systems where $N$ is small~\cite{zeb2018exact}.\\
The shift between the zeroth and first-order polaritons, corresponding to the energy difference of the ground state vibrational levels, provides information analogous to a Raman spectrum. A discrepancy in the Rabi splitting, $\Delta = 2\lambda(\sqrt{N} - \sqrt{N-1})$, causes a blue shift in the first-order peaks (see Fig.~\ref{fig:results}a inset), due to the decrease in $\lambda\sqrt{N-1}$ as $N$ decreases while $\lambda\sqrt{N}$ remains constant. The blue-shift $\Delta$ scales as $1/\sqrt{N}$, and the peak height as $1/N$; this latter scaling can be understood from Eq. 8 since each action of $\bm{v}_0$ and $\bm{v}_0^\dagger$ is proportional to $N^{-1/2}$. \\

Even though we cannot provide an expression for $d_{N,k}$ when $k\ge2$, the truncation of Eq.~\ref{Eq:contd_frac_full} at $\bm{H}_{e,k}$ gives $D^R_N(\omega)$ up to $\mathcal{O}(N^{-k})$. To best illustrate the spectroscopic implications of the $\mathcal{O}(N^{-2})$ corrections to $d_{N,0}(\omega)$, we consider a cavity coupled to $N_A$ and $N_B$ molecules of two different species. Eq.~\ref{bosonic_Hamiltonian} can be trivially generalized to account for two sets of bosonic modes corresponding to the different molecules~\cite{perez2023frequency}. Each molecule is modeled as a 3-level system with different vibrational gaps, $\omega_{v,A}$ and $\omega_{v,B}$. For simplicity, let the electronic transition frequencies, $\omega_{e,\varphi^{(e)}_{0,i}}-\omega_{g,\varphi^{(g)}_{0,i}}$, is identical for both species $i=A,B$. Fig.~\ref{fig:results}b shows $A(\omega)$ for this setup. Including the second-order correction introduces sidebands shifted from the zeroth-order polaritons by $2\omega_{g,A}$, $2\omega_{g,B}$, and $\omega_{g,A}+\omega_{g,B}$ (see the three peaks appearing only in blue in the inset) due to a Raman process involving the second-order polaritons. These shifts correspond to the sum of Raman transition frequencies created in two molecules of the same or different species, in a similar way that entangled photons induce collective resonances between different molecules~\cite{Muthukrishnan2004inducing, Seidelmann2022two}.\\

\noindent \textbf{Feynman pathways}.\textemdash  \begin{figure*}
\begin{centering}
\includegraphics[width=0.9\linewidth]{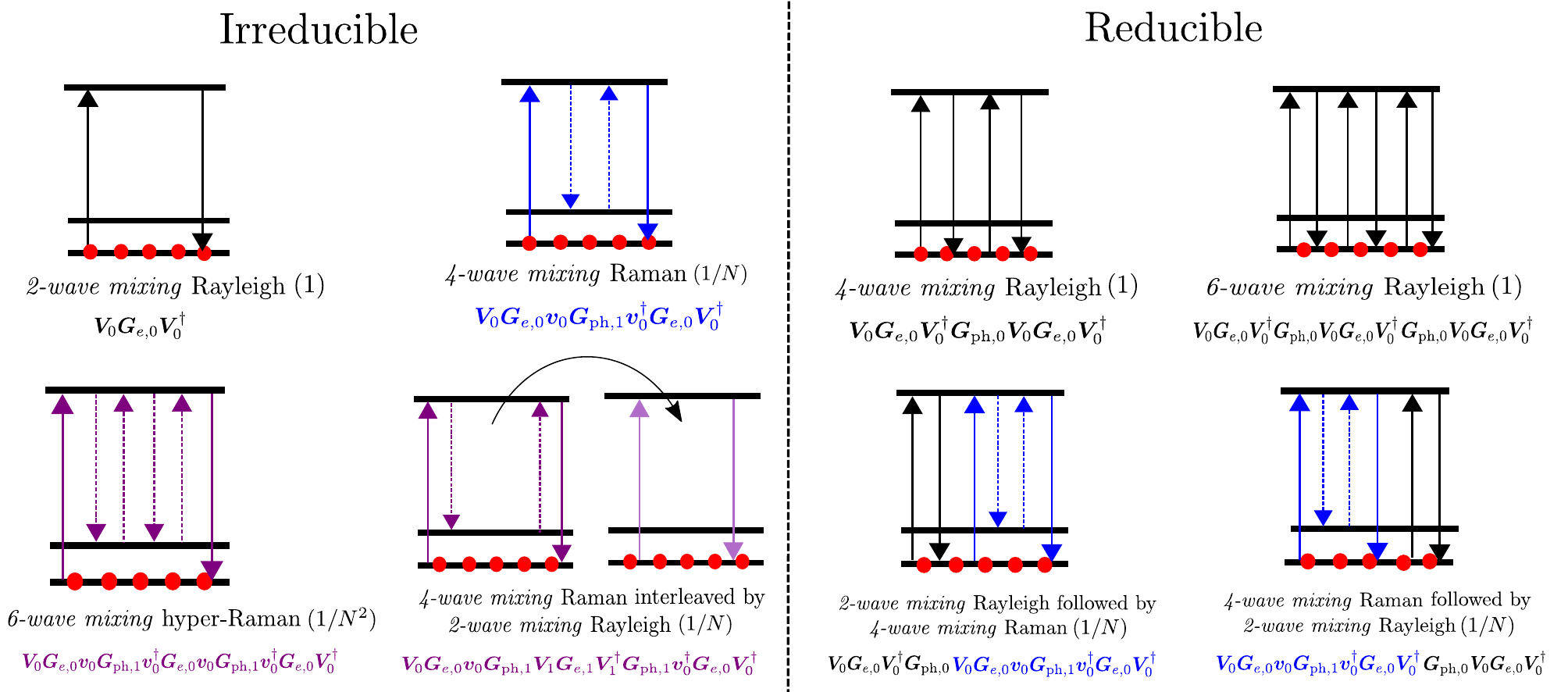}
\par\end{centering}
\caption{\emph{Cavity-mediated nonlinear molecular processes}: Ladder diagrams for an ensemble of identical three-level systems with two ground-state vibrational levels and one excited state, showing second, fourth, and sixth-order transitions mediated by the cavity, classified into \emph{reducible} and \emph{irreducible} nonlinear susceptibilities with their correction order in the parentheses. The \textbf{second-order} term represents a \emph{2-wave mixing} collective Rayleigh process. The \textbf{fourth-order} term includes two contributions: a collective \emph{4-wave mixing} Rayleigh and a \emph{4-wave mixing} Raman. The \textbf{sixth-order} term has five contributions: a collective \emph{6-wave mixing} Rayleigh, two \emph{reducible} terms involving \emph{2-wave mixing} Rayleigh and \emph{4-wave mixing} Raman, a \emph{6-wave mixing} Raman, and a cascade process involving two molecules with a \emph{4-wave mixing} Raman interleaved by a \emph{2-wave mixing} Rayleigh. The distinction between \emph{reducible} and \emph{irreducible} contributions lies in whether the excitation, after the initial $\bm{V}_0$ interaction with the molecular ensemble, fully returns to $\bm{H}_{\text{ph},0}$ $(\bm{G}_{\text{ph},0})$ via the $\bm{V}_0^\dagger$ interaction at any intermediate step.
\label{fig:diagram_2}}
\end{figure*} For simplicity, we return to the discussion of an ensemble of identical molecules. It turns out that Eq.~\ref{Eq:contd_frac_full} can be rewritten as [Supplemental Material: Sec. \textcolor{blue}{S5}],
\begin{widetext}
  \begin{equation}\label{eq. D_R_general}
    D^{R}_N(\omega)=\frac{1}{\omega-\omega_{\text{ph}}+i\kappa/2+\sum_{l=0}^{\infty}\big(\frac{\omega_{\text{ph}}}{2}\big)^l\chi_N^{(2l+1)}(\{\omega,\omega-\omega_{\text{ph}}\}^l,\omega)},
\end{equation}  
\end{widetext}
\noindent where $\sum_{l=0}^{\infty}\big(\frac{\omega_{\text{ph}}}{2}\big)^l\chi_N^{(2l+1)}(\{\omega,\omega-\omega_{\text{ph}}\}^l,\omega)$ is the sum over all the \emph{irreducible} Rayleigh and Raman nonlinear susceptibilities of the $N$-molecule ensemble ($\{\omega,\omega-\omega_{\text{ph}}\}^l$ denotes a string where $\omega$, $\omega-\omega_{\text{ph}}$ are repeated $l$ times). We shall now define the concept of \emph{reducible} (\emph{irreducible}) nonlinear susceptibilities, which describe processes that can (cannot) be decomposed into products of other bare susceptibilities, $\chi^{2l+1}_N(\sum_{l=1}^{2l+1}\omega_{l},\dots,\omega_1)$~\cite{MukamelBook, glenn2015photon}; as far as we aware, this concept does not exist in the literature, although similar ideas are standard in diagrammatic many-body theory~\cite{fetter2003quantum}. The key difference between \emph{reducible} and \emph{irreducible} susceptibilities lies in whether, upon initial excitation of molecules via $\bm{V}_0^{\dagger}$, the excitation fully returns to the cavity mode via $\bm{V}_0$ at an intermediate step or not; see Fig.~\ref{fig:diagram_2}. It is easy to see that the expansion of the self-energy in Eq.~\ref{Eq:contd_frac_full} into terms of different order in light-matter coupling gives rise to irreducible diagrams only ($\bm{V}_0$ and $\bm{V}_0^\dagger$ act only at the beginning and end). Another interesting observation is that interpreting Eq.~\ref{eq. D_R_general} as a geometric series gives rise to all the possible \emph{reducible} and {irreducible} diagrams for cavity linear response.
\\
\noindent \textbf{Experimental considerations.}\textemdash  The spectral resolution of the polaritonic linear response is mainly determined by its cavity lifetime, $2\pi/\kappa$. 
Given that the peak heights correspond to the vacuum-induced nonlinearities determined by the single-molecule light-matter coupling $\lambda$, $\kappa$ must be comparable to $\lambda$ to resolve these features. This insight also helps understand why, in typical experiments with low-Q cavities, $\kappa\gg \lambda$, the truncation of the $1/N$ expansion at $d_{N,0}$ works so well. Another reason why these nonlinearities are not typically observed is due to the use of multimode cavities, given that the here-predicted first-order polariton peaks will instead blur into broad continua (the Stokes Raman photon can be emitted into any cavity mode)\cite{cao2023polariton}. Thus, effective single-mode cavities are needed, which, from an experimental standpoint, translate into cavities with large free spectral ranges or with dispersionless cavities.~\cite{Xue2016dispersionless,Fujii2020dispersion,Jomaso2024intracavity,Baboux2016flatband}. \\
\\
\noindent \textbf{Summary and conclusions.}\textemdash We demonstrated that while the thermodynamic limit of molecular polaritons in typical Fabry-Perot cavities, with broad linewidths, is well-captured by classical linear optics, high-Q, single-mode cavities can reveal the otherwise hidden nonlinearities expected in systems under SC. To capture these effects, we derived a general expression for the linear response of polaritons beyond the classical regime, expressed in terms of \emph{irreducible} Rayleigh and Raman nonlinear susceptibilities of the molecular ensemble. By leveraging the timescale hierarchy in the polariton problem, we presented a $1/N$ expansion of this expression. While we focused on the spectroscopic aspects of the problem, we believe these higher-order coherent processes in the molecular ensemble could serve as a pathway for harnessing quantum resources such as entanglement and nontrivial photon statistics, which will be explored in future works.

\begin{acknowledgments}
A.K. thanks Sricharan Raghavan-Chitra, Juan B. P\'erez-S\'anchez, Piper Fowler-Wright, Juan Carlos Obeso Jureidini, and Kai Schwennicke for useful discussions. J.Y.Z. and A.K. thank Abraham Nitzan for providing the connection with Raman and Rayleigh scattering and for asking whether the result holds for multimode cavities.
\end{acknowledgments}

\bibliographystyle{apsrev4-1-custom}
\bibliography{linear_response}

\end{document}


\preprint{AIP/123-QED}

\title{Hidden nonlinear optical susceptibilities in linear polaritonic spectra: Supplemental material}

\author{Arghadip Koner}
\author{Joel Yuen-Zhou}%
 \email{joelyuen@ucsd.edu}
\affiliation{%
 Department of Chemistry and Biochemistry, University of California San Diego, La Jolla, California 92093
}%
\date{\today}

\maketitle
\section{The Hamiltonian}\label{sec:Hamiltonian_supp}
\noindent The SC setup consisting of a photonic cavity with a single mode of frequency $\omega_{\text{ph}}$ coupled to $N$ non-interacting molecules is typically modeled using the Holstein-Tavis-Cummings Hamiltonian  extended to arbitrary vibronic structures~\cite{perez2024cute}, 
\begin{align}\label{eq:htc}
{H_{HTC}}&=\hbar\omega_{{\text{ph}}}{a}^{\dagger}{a}+\sum_{i=1}^{N}\left(\hat{T}_{i}+V_{g}(q_{i})|g_{i}\rangle\langle g_{i}|+V_{e}(q_{i})|e_{i}\rangle\langle e_{i}|\right)\nonumber\\
&+\hbar\lambda\sum_{i}^{N}\left(|e_{i}\rangle\langle g_{i}|{a}+|g_{i}\rangle\langle e_{i}|{a}^{\dagger}\right),
\end{align} where $\hat{T}$ is the kinetic energy operator, $V_{g/e}$ are the ground/excited potential energy surfaces (PES), $\hbar\lambda$ is the single-molecule light-matter coupling strength, and ${a}$ is the annihilation operator of a photon in the cavity mode. 
P\'erez-S\'anchez \emph{et. al.} has shown that starting from a permutationally invariant initial state and exploiting the symmetries under the permutations of the molecules, the setup obeys a second-quantized bosonic Hamiltonian~\cite{perez2024bosons,Pizzi2023light,biedenharn1965quantum, perez2024cute}, 
\begin{eqnarray}\label{bosonic_Hamiltonian}
   \nonumber H &=& H_{0} + V\\   
   \nonumber &=&\bigg[\hbar\omega_{\text{ph}}a^\dagger a + \hbar \sum_{j=1}^{M_g} \omega_{g,j} b_j^\dagger b_j + \hbar \sum_{j=1}^{M_e} \omega_{e,j} B_j^\dagger B_j\bigg]+ \\  
    & &  \bigg[-\hbar \lambda \sum_{jj'} \langle \varphi^{(e)}_{j'}| \varphi^{(g)}_{j} \rangle ab_j B_{j'}^\dagger - \text{h.c.}\bigg].
\end{eqnarray}
 \noindent Here, $H_0$ represents the zeroth-order contribution from the bare cavity and the molecules, and $V$ models the cavity-molecule interaction with coupling strength $\hbar \lambda$. The operators $b_j$ and $B_{j'}$ annihilate a molecule in a vibronic state $|g, \varphi^{(g)}_{j}\rangle$ (the first and the second index represents the electronic state and the vibrational states, respectively), and a molecule in the vibronic state $|e, \varphi^{(e)}_{j'}\rangle$, respectively. The eigenstates of the non-interacting part of the Hamiltonian, $H^{0}$, are represented as $|n_{\text{ph}}; n_1,n_2,\dots,n_{M_g};n'_1,n'_2,\dots,n'_{M_e}\rangle$, where $M_g$ and $M_e$ are the sizes of the vibrational bases used in the model, $n_i(')$ is the number of molecules in the $i^{\text{th}}$ vibrational level of the ground (excited) electronic state, and $n_{\text{ph}}$ is the number of photonic excitations~\cite{Philpott1971theory,spano2011vibronic}. \noindent In the rest of this work, we will consider our initial state to be $|0;N,\vec{0};\vec{0}\rangle$, which is a photonless state with all the molecules in the global ground state (zero temperature). Although the bosonic picture is valid for an arbitrary number of excitations in the system, the linear response regime restricts us to the first excitation manifold in the Hamiltonian. We have previously shown in Ref.~\cite{perez2024cute} that in the eigenbasis of $H_{0}$, $H$ admits a block tridiagonal representation:
 \begin{equation}\label{CUT-E_matrix}
H_1=\begin{pmatrix}
\bm{H}_{\text{ph},0} & \bm{V}_{0} & 0 & \dots & 0 & 0 & 0 \\
\bm{V}_{0}^{\dagger} & \bm{H}_{e,0} & \bm{v}_{0} & \dots & 0 & 0 &  \\
0 & \bm{v}_{0}^{\dagger} & \bm{H}_{\text{ph},1} & \bm{V}_{1} & 0 & 0 & 0 \\
\vdots & \vdots & \vdots & \ddots & \vdots & \vdots & \vdots \\
0 & 0 & 0 & \dots & \bm{V}_{N-1}^{\dagger} & \bm{H}_{e,N-1} & \bm{v}_{N-1} \\
\vdots & \vdots & \vdots & \vdots & 0 & \bm{v}_{N-1}^{\dagger} & \bm{H}_{\text{ph},N}.
\end{pmatrix}
\end{equation}
\noindent The sub-blocks of $H_1$ of the form $\begin{pmatrix}
\bm{H}_{\text{ph},n} & \bm{V}_{n}  \\
\bm{V}_{n}^{\dagger} & \bm{H}_{e,n}\end{pmatrix}$ have $n$, the number of molecules with electronic ground state (GS) phonons as a conserved quantity, and only the slow $\bm{v}_n\propto \lambda$ interactions can change the number of molecules with electronic GS phonons, leading to a timescale separation in the quantum dynamics. The block tridiagonal structure of $H_1$ allows for its representation as a nearest neighbor coupled chain model~\cite{butler1973self,
pastawskii1983matrix, pastawski1997nmr,
Martinez2003floquet,hangii2018book,
Hangii1978continued, Duneczky1988lanczos,Yang1989block},
\begin{widetext}    
\begin{equation}\label{CUT-E_diagram} 
{\boxed{\bm{H}_{\text{ph},0}\xleftrightarrow[\bm{V}_{0}]{\bm{V}_{0}^{\dagger}}\bm{H}_{e,0}}}\xleftrightarrow[\bm{v}_{0}]{\bm{v}_{0}^{\dagger}}{\boxed{\bm{H}_{\text{ph},1}\xleftrightarrow[\bm{V}_{1}]{\bm{V}_{1}^{\dagger}}\bm{H}_{e,1}}}\xleftrightarrow[\bm{v}_{1}]{\bm{v}_{1}^{\dagger}}\dots\boxed{\bm{H}_{\text{ph},N-1}\xleftrightarrow[\bm{V}_{N-1}]{\bm{V}_{N-1}^{\dagger}}\bm{H}_{e,N-1}}\xleftrightarrow[\bm{v}_{N}]{\bm{v}_{n}^{\dagger}}\bm{H}_{\text{ph},N}.
\end{equation}
\end{widetext}
\noindent We call this the CUT-E (collective dynamics using truncated equations) diagram, based on Ref.~\cite{perez2023simulating}, where this timescale separation was first formulated. We will use it to develop a diagrammatic approach later. 
\section{The photon Green's function}\label{Sec:photonGF_supp}

\begin{figure}
\begin{centering}
\includegraphics[width=\linewidth]{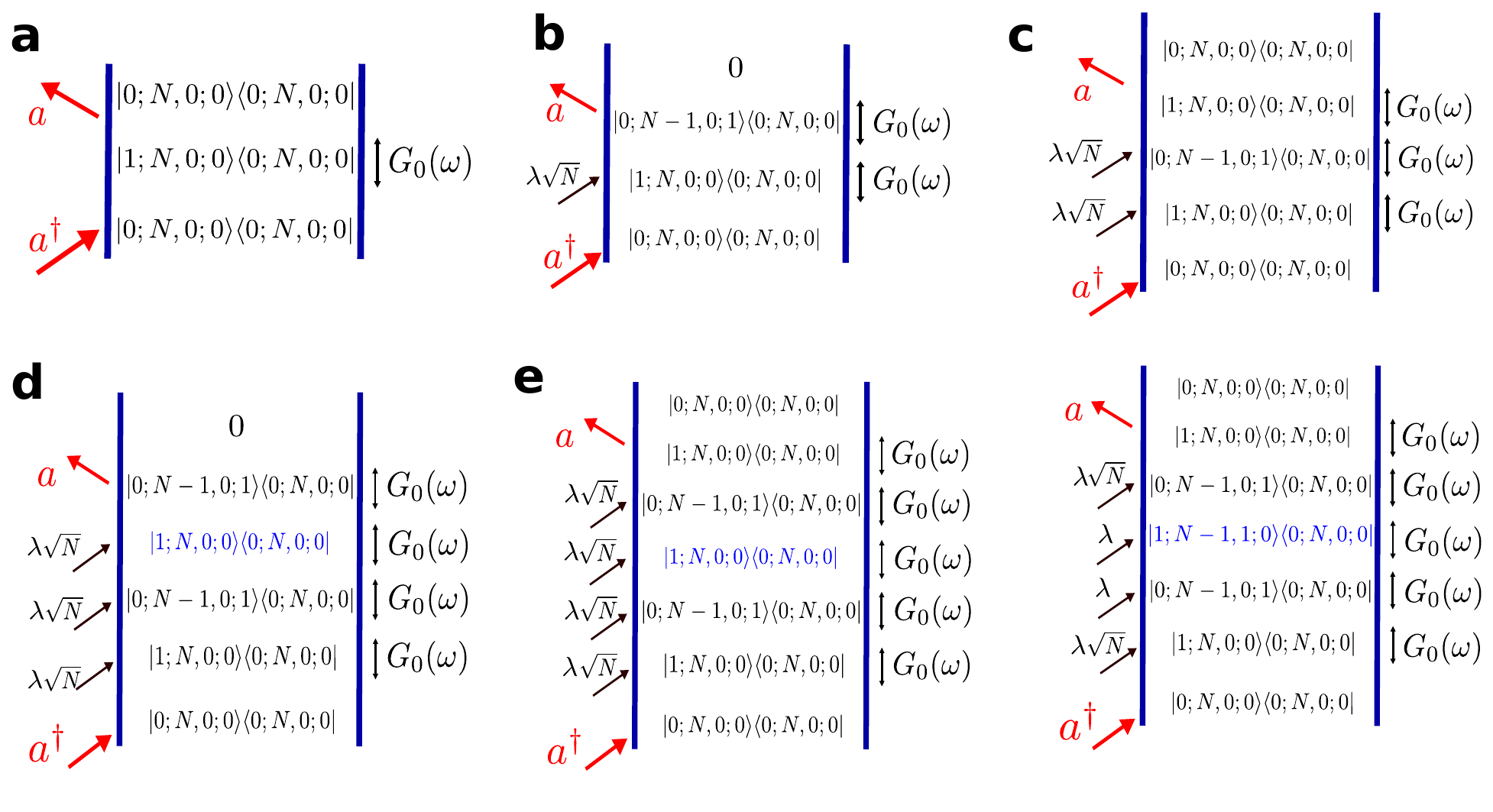}
\par\end{centering}
\caption{The \emph{ket}-only DS-FDs for the a. zeroth, b. first, c. second, d. third, and e. fourth order. Notice that only the even orders have non-zero contributions. It can be checked that for the $\mathcal{T}=0$ photonless initial state, at a given order, these are the only diagrams that contribute. In the fourth order, we notice that the second diagram involving a Raman process is penalized by a factor of $1/N$. \label{fig:dsfd_ket}}
\end{figure}

\noindent The linear spectroscopic observables for the polaritonic setup with $N$ molecules are functions of the photon Green's function, $D^{R}_N(\omega)=-i\int_{-\infty}^{\infty}dte^{i\omega t}\Theta(t)\langle [a(t),a^{\dagger}]\rangle$~\cite{yuen2023linear,cwik2016excitonic}; the average $\langle \dots \rangle$ is computed with respect to the initial state. We have the transmission, $T(\omega)$, absorption, $A(\omega)$, and reflection, $R(\omega)$, 
\begin{subequations}\label{eq:final_formulas}
\begin{align}
T(\omega) & =(\kappa^2/{4})|D^{R}_N(\omega)|^{2},\label{eq:T_w}\\
R(\omega) & =1+\kappa\Im D^{R}_N(\omega)+(\kappa^{2}/4)|D^{R}_N(\omega)|^{2},\label{eq:R_w}\\
A(\omega) & =-(\kappa/2)[\kappa|D^{R}_N(\omega)|^{2}+2\Im D^{R}_N(\omega)].\label{eq:A_w}
\end{align}
\end{subequations}
For our zero-temperature ($\mathcal{T}=0$) photonless initial state, we can eliminate one of the terms in the commutator, giving 
\begin{eqnarray}
    D^{R}_N(\omega)|_{\mathcal{T}=0}&=&-i\int_{-\infty}^{\infty}dte^{i\omega t}\Theta(t)\langle a e^{-iHt}a^{\dagger}\rangle\\
    &=& \langle a G(\omega) a^{\dagger}\rangle,
\end{eqnarray} 
where $G(\omega)=-i\int_{-\infty}^{\infty}dte^{i\omega t}\Theta(t) e^{-iHt}=\frac{1}{\omega-H+i0^+}$ is the full system's Green's function. Now to compute $D^R_N(\omega)$, we can use the Dyson series, 
\begin{eqnarray}
    \nonumber G&=&G_0+G_0VG\\
    &=& G_0+G_0VG_0+G_0VG_0VG_0+\dots    
\end{eqnarray} 
    
\noindent where $G_0=\frac{1}{\omega-H_0+i0^+}$ is the Green's function of the noninteracting part. Typically double-sided Feynmann diagrams (DS-FDs)~\cite{MukamelBook} are used in the theory of nonlinear spectroscopy to compute the terms in the Dyson series. \textbf{However, notice that elimination of one of the terms of the commutator at $\mathcal{T}=0$ implies that exclusively the \emph{ket}-only DS-FDs have non-zero contributions to $D^{R}_N(\omega)$}. The physical implication of this result is that only optical coherences contribute to the response in the linear regime, a result that is also consistent with the understanding of linear response outside of cavities. In Fig~\ref{fig:dsfd_ket}, we show the \emph{ket}-only diagrams up to the fourth order. Notice that only the terms, even orders in $V$, contribute to a non-zero response. This is a consequence of the fact that the interaction $V$ can only exchange the excitation between the cavity and the molecules.\\

\noindent The result that only \emph{ket}-only diagrams contribute allows us to use the CUT-E diagram instead of the DS-FDs to compute the different orders in the Dyson series. In addition to the notational simplicity, the CUT-E diagram separates the nonlinear transitions based on their timescales. This will later help us obtain the $1/N$ expansion. Here, we state the rules for using the CUT-E diagram to compute the terms of the Dyson series:

\begin{enumerate}\label{rules}
    \item The $(2m)^\text{th}$ order term in $D^R_N(\omega)$ consists of all the $2m$ step paths in the CUT-E diagram \emph{starting from} and \emph{returning} to $\bm{H}_{\text{ph},0}$; the order here is calculated with respect to the light-matter interaction, $V$,
    
    \item Every time we jump from one box to another in the CUT-E diagram, we encounter a Raman process mediated by a $\bm{v}^{(\dagger)}_{k}$ which are penalized by a factor of $1/\sqrt{N}$. 
\end{enumerate}
It can be seen (and proven explicitly using mathematical induction) that the CUT-E diagram with the aforementioned rules is equivalent to computing \emph{ket}-only diagrams. Below, we will show a few examples of using the CUT-E diagram to compute the terms of $D^{R}(\omega)$ at different orders of the Dyson series with the $k^\text{th}$ order term denoted as $D^{R,(k)}(\omega)$. 
\begin{enumerate}
    \item \textbf{Zeroth order term}: 0 step paths. We have $D^{R,(0)}(\omega)=\bm{G}_{\text{ph},0}(\omega)=\frac{1}{\omega-\omega_{	\text{ph}}+i\kappa/2}$.
    \item \textbf{Second order term}: We want to count the $2$ step paths. We have just one
of them: $\boxed{\bm{H}_{\text{ph},0}\xleftrightarrow[2]{1}\bm{H}_{e,0}}$ (the numbers on the top (bottom) represent the counting of the step forward (backward)). Thus, second-order photon Green's function, $D^{R,(2)}(\omega)$, is then given
as 
\begin{eqnarray}\label{eq:second}
    D^{R,(2)}(\omega)&=&\bm{G}_{\text{ph},0}(\bm{V}_0 \bm{G}_{e,0} \bm{V}^\dagger_0)\bm{G}_{\text{ph},0},
\end{eqnarray}
\item \textbf{Fourth order term:}  All the 4 step paths are presented in the table below . The fourth-order photon Green's function, $D^{R,(4)}(\omega)$, is then given as the sum of all the Dyson series terms, 

\begin{center}
  \begin{tabular}{|c|c|c|}
\hline 
$D_{N}^{R,(4)}(\omega)$ & Pathways in CUT-E diagram & Dyson series term\tabularnewline
\hline 
 & $\boxed{\bm{H}_{\text{ph},0}\xleftrightarrow[2,4]{1,3}\bm{H}_{e,0}}=2\cdot(\bm{H}_{\text{ph},0}\leftrightarrow\bm{H}_{e,0})$ & $\bm{G}_{\text{ph},0}\cdot\big(\bm{V}_0\bm{G}_{e,0}\bm{V}_0^\dagger \bm{G}_{\text{ph},0}\big)^2$\tabularnewline
\hline 
 & $\boxed{\bm{H}_{\text{ph},0}\xleftrightarrow[4]{1}\bm{H}_{e,0}}\xleftrightarrow[3]{2}\bm{H}_{\text{ph},1}$ & $\bm{G}_{\text{ph},0}\big(\bm{V}_0\bm{G}_{e,0}\bm{v}_0 \bm{G}_{\text{ph},1}\bm{v}_0^\dagger \bm{G}_{e,0}\bm{V}_0^\dagger\big)\bm{G}_{\text{ph},0}$\tabularnewline
\hline 
\end{tabular} 
\end{center}

\item \textbf{Sixth order term:} We want all the 6-step paths in the diagram. We can have five of them presented in the table below. Note that $D^{R,(6)}_N(\omega)$ is the sum of all the Dyson series terms.

\begin{center}
    \begin{tabular}{|c|c|c|}
        \hline
        \( D^{R,(6)}_N(\omega) \) & Pathways in CUT-E diagram  &Dyson series term \\ \hline
        & \(\bm{H}_{\text{ph},0} \xleftrightarrow[2,4,6]{1,3,5} \bm{H}_{e,0} = 3 \cdot (\bm{H}_{\text{ph},0} \leftrightarrow \bm{H}_{e,0})\) & \(\bm{G}_{\text{ph},0} \left(\bm{V}_0 \bm{G}_{e,0} \bm{V}_0^\dagger \bm{G}_{\text{ph},0}\right)^3\) \\ \hline
        & \(\boxed{\bm{H}_{\text{ph},0} \xleftrightarrow[2,6]{1,3} \bm{H}_{e,0}} \overset{4}{\underset{5}{\leftrightarrow}} \bm{H}_{\text{ph},1}\) & \(\bm{G}_{\text{ph,0}}(\bm{V}_{0} \bm{G}_{e,0} \bm{V}_{0}^{\dagger}) \bm{G}_{\text{ph,0}} (\bm{V}_{0} \bm{G}_{e,0} \bm{v}_{0} \bm{G}_{\text{ph},1} \bm{v}_{0}^{\dagger} \bm{G}_{e,0} \bm{V}_{0}^{\dagger}) \bm{G}_{\text{ph,0}}\) \\ \hline
        & \(\boxed{\bm{H}_{\text{ph},0} \xleftrightarrow[4,6]{1,5} \bm{H}_{e,0}} \overset{2}{\underset{3}{\leftrightarrow}} \bm{H}_{\text{ph},1}\) & \(\bm{G}_{\text{ph,0}} (\bm{V}_{0} \bm{G}_{e,0} \bm{v}_{0} \bm{G}_{\text{ph},1} \bm{v}_{0}^{\dagger} \bm{G}_{e,0} \bm{V}_{0}^{\dagger}) \bm{G}_{\text{ph,0}} (\bm{V}_{0} \bm{G}_{e,0} \bm{V}_{0}^{\dagger}) \bm{G}_{\text{ph,0}}\) \\ \hline
        & \(\boxed{\bm{H}_{\text{ph},0} \overset{1}{\underset{6}{\leftrightarrow}} \bm{H}_{e,0}} \xleftrightarrow[3,5]{2,4} \bm{H}_{\text{ph},1}\) & \(\bm{G}_{\text{ph,0}} (\bm{V}_{0} \bm{G}_{e,0} \bm{v}_{0} \bm{G}_{\text{ph},1} \bm{v}_{0}^{\dagger} \bm{G}_{e,0} \bm{v}_{0} \bm{G}_{\text{ph},1} \bm{v}_{0}^{\dagger} \bm{G}_{e,0} \bm{V}_{0}^{\dagger}) \bm{G}_{\text{ph,0}}\) \\ \hline
        & \(\boxed{\bm{H}_{\text{ph},0} \xleftrightarrow[6]{1} \bm{H}_{e,0}} \xleftrightarrow[5]{2} \boxed{\bm{H}_{\text{ph},1} \xleftrightarrow[4]{3} \bm{H}_{e,0}}\) & \(\bm{G}_{\text{ph,0}} (\bm{V}_{0} \bm{G}_{e,0} \bm{v}_{0} \bm{G}_{\text{ph,1}} \bm{V}_{1} \bm{G}_{e,1} \bm{V}_{1}^{\dagger} \bm{G}_{\text{ph,1}} \bm{v}_{0}^{\dagger} \bm{G}_{e,0} \bm{V}_{0}^{\dagger}) \bm{G}_{\text{ph,0}}\) \\ \hline
    \end{tabular}
\end{center}
 
\end{enumerate}

\noindent We will come to these results in Sec.~\ref{sec:FD-1} where we discuss \emph{reducible} and \emph{irreducible} diagrams. 

\section{Some results for matrix Green's function and matrices}\label{sec.matrix_gf}
\noindent This section is quite standard but is presented for completeness and for pedagogical purposes. ~\cite{pastawski1997nmr,Velev2004gf,pastawskii1983matrix}. 
\subsection{Green's function of two connected systems~\label{Green_func_supp}} 

\noindent Let us first consider two uncoupled systems, $A$ and $B$ with their bare Green's functions $G_{AA}^{0}$ and $G_{BB}^{0}$, respectively. If these systems are coupled by an interaction $V=V_{AB}+V_{BA}$,

\[
G=(E-H_{0}-V)^{-1}=((G_{0})^{-1}-V)^{-1}.
\]

\noindent Explicitly,

\[
G=\begin{pmatrix}(G_{AA}^{0})^{-1} & -V_{AB}\\
-V_{BA} & (G_{BB}^{0})^{-1}
\end{pmatrix}^{-1}.
\]

\noindent To find the inverse, we will use the method of Schur complement
(see sec. \ref{subsec:Schur-complement}). We have

\begin{align*}
G_{AA} & =((G_{AA}^{0})^{-1}-V_{AB}G_{BB}^{0}V_{BA})^{-1},\\
G_{BB} & =((G_{BB}^{0})^{-1}-V_{BA}G_{AA}^{0}V_{AB})^{-1},\\
G_{AB} & =-G_{AA}^{0}(-V_{AB})((G_{BB}^{0})^{-1}-V_{BA}G_{AA}^{0}V_{AB})^{-1}\\
 & =G_{AA}^{0}V_{AB}G_{BB},\\
G_{BA} & =G_{BB}^{0}V_{BA}G_{AA}.
\end{align*}

\subsection{Schur complement\label{subsec:Schur-complement}}

\noindent Let $A,B,C,D$ be $p\times p,p\times q,q\times p$ and $q\times q$ matrices of complex numbers. and construct,

\[
M=\begin{pmatrix}A & B\\
C & D
\end{pmatrix},
\]

\noindent as a $(p+q)\times(p+q)$ matrix. If $D$ is invertible, the Schur
complement of the block $D$ in $M$ is the $p\times p$
matrix,

\[
M/D:=A-BD^{-1}C.
\]

\noindent Similarly, if $A$ is invertible, the Schur complement of block
$A$ is the $q\times q$ matrix,

\[
M/A:=D-CA^{-1}B.
\]

\noindent Consequently, the inverse of $M$ can be written in terms of the
Schur complements as,

\begin{align*}
M^{-1} & =\begin{pmatrix}A & B\\
C & D
\end{pmatrix}^{-1}\\
 & =\begin{pmatrix}(M/D)^{-1} & -A^{-1}B(M/A)^{-1}\\
-D^{-1}C(M/D)^{-1} & (M/A)^{-1}
\end{pmatrix}.
\end{align*}
\subsection{Series expansion of the matrix inverse\label{subsec:Series-expansion}}

\noindent Given a matrix $M=(A-B)$. We can express its inverse as,

\begin{align}\label{eq:matrix_series}
\nonumber M^{-1} & =(A-B)^{-1}\\
 \nonumber & =A^{-1}+A^{-1}B(\bm{I}-A^{-1}B)^{-1}A^{-1}\\
 \nonumber & =A^{-1}+A^{-1}B(\bm{I}+A^{-1}B+A^{-1}BA^{-1}B+\dots)A^{-1}\\
 \nonumber & =A^{-1}+A^{-1}BA^{-1}+A^{-1}BA^{-1}BA^{-1}+A^{-1}BA^{-1}BA^{-1}BA^{-1}+\dots\\
 & =A^{-1}(\bm{I}+BA^{-1}+(BA^{-1})^{2}+\dots)
\end{align}

\noindent Here, from the first to the second step, we have used the Woodbury
matrix identity: $(A+UCV)^{-1}=A^{-1}-A^{-1}U(\bm{I}+VA^{-1}U)^{-1}VA^{-1}$
with $U=-B,C=\bm{I},$ and $V=\bm{I}$. In the second to the
third line, we have used the Neumann expansion: $(\bm{I}-A^{-1}B)^{-1}=\bm{I}+A^{-1}B+A^{-1}BA^{-1}B+\dots$.

\section{Matrix continued fraction expansion}\label{sec:matrix_contd_frac}
\noindent We are interested in the Green's function of $H_1$ (Eq.~\ref{CUT-E_matrix}) projected onto $\bm{H}_{\text{ph},0}$. We can now use the standard techniques of matrix Green's functions for connected systems as presented in Sec.~\ref{sec.matrix_gf}. Writing $H_1$ as,
\begin{equation}
    H_1=\begin{pmatrix}\bm{H}_{\text{ph},0} & \bm{V}_{0}\\
\bm{V}_{0}^{\dagger} & \tilde{\bm{H}}_{e,0}
\end{pmatrix},
\end{equation}
\noindent we have the photon Green's function, $D^{R}_N(\omega)$, as 
\begin{eqnarray}\label{DRw1}
    D^{R}_N(\omega)&=& \frac{1}{\omega-\omega_{\text{ph}}+i\kappa/2-\Sigma_{e,0}}
\end{eqnarray}
\noindent where 
\begin{eqnarray}
    \nonumber \Sigma_{e,0}&=&\bm{V}_{0}(\omega-\tilde{\bm{H}}_{e,0}+i\gamma/2)^{-1}\bm{V}_{0}^{\dagger},
\end{eqnarray}
is the cavity self-energy due to its coupling to the $\tilde{\bm{H}}_{e,0}$ block. Now we can write $\tilde{\bm{H}}_{e,0}$ as,
\begin{equation}  \tilde{\bm{H}}_{e,0}=\begin{pmatrix}\bm{H}_{e,0} & \bm{v}_{0}\\
\bm{v}_{0}^{\dagger} & \tilde{\bm{H}}_{\text{ph},1}.
\end{pmatrix}.
\end{equation}
Using this, we can obtain $\Sigma_{e,0}$, in terms of $\bm{\Sigma}_{\text{ph},1}$.
\begin{equation}
   \Sigma_{e,0} = \bm{V}_{0}\big(\omega-\bm{H}_{e,0}+i\gamma/2-\bm{\Sigma}_{\text{ph},1}\big)^{-1}\bm{V}_{0}^{\dagger}.
\end{equation}
\noindent Again $\bm{\Sigma}_{\text{ph},1}=\bm{v}_{0}\big(\omega-\bm{H}_{\text{ph},1}+i\kappa/2-\bm{\Sigma}_{e,1}\big)^{-1}\bm{v}_{0}^{\dagger}$ is obtained in terms of the self-energy of the next step, $\bm{\Sigma}_{e,1}$. Proceeding this way, we get a recursive algorithm, where the self-energy at the $k^{\text{th}}$ step in the matrix depends on the $(k+1)^\text{th}$ step. Thus we have,
\begin{subequations}
\begin{align}
\bm{\Sigma}_{e,k}&=\bm{V}_{k}\big(\omega-\bm{H}_{e,k}+i\gamma/2-\bm{\Sigma}_{\text{ph},k+1}\big)^{-1}\bm{V}_{k}^{\dagger},\\
    \bm{\Sigma}_{\text{ph},k+1}&=\bm{v}_{k}\big(\omega-\bm{H}_{\text{ph},k+1}+i\kappa/2-\bm{\Sigma}_{e,k+1}\big)^{-1}\bm{v}_{k}^{\dagger},
\end{align}
\end{subequations}
\noindent as the recursion relations for $0\le k<N$. For finite $N$, the series truncates at $\bm{\Sigma}_{\text{ph},N}=\bm{v}_{N-1}\big(\omega-\bm{H}_{\text{ph},N}+i\kappa/2\big)^{-1}\bm{v}_{N-1}^{\dagger}$. 
\section{{Reducible} and {irreducible} Feynman diagrams}\label{sec:FD-1}

\noindent Defining
\begin{equation}\label{lorentzian}
    \mathcal{I}(\omega',\Gamma)=\frac{1}{\omega-\omega'+i\Gamma/2},
\end{equation}
\noindent we have the explicit expressions for the  `Rayleigh' and `Raman' nonlinear susceptibilities of the molecules, $\chi^{2l+1}_N(\sum_{l=1}^{2l+1}\omega_{l},\dots,\omega_1)$ ~\cite{MukamelBook} as, 

\begin{subequations}\label{eq:susceptibilities}
\begin{align}
   \bigg(\frac{\omega_{\text{ph}}}{2}\bigg)\chi_N^{(1)}(\omega) &= -\bm{V}_0 \bm{G}_{e,0} \bm{V}^\dagger_0 \notag \\    
    &= -\sum_{m_{e}}\mu_{0_{g}m_{e}}\lambda\sqrt{N}\mathcal{I}(\omega_{e}+m_{e},\gamma)\lambda\sqrt{N}\mu_{m_{e}0_{g}}, \\ 
     \bigg(\frac{\omega_{\text{ph}}}{2}\bigg)^{2}\chi_N^{(3)}(\omega,\omega-\omega_{\text{ph}},\omega) &= -\bm{V}_0\bm{G}_{e,0}\bm{v}_0 \bm{G}_{\text{ph},1}\bm{v}_0^\dagger \bm{G}_{e,0}\bm{V}_0^\dagger \notag \\     
    &= -\sum_{m_{e}, m'_{e}, m_{g}} \lambda\sqrt{N}\mu_{0_{g}m'_{e}} \, \mathcal{I}(\omega_{e}+m'_{e},\gamma) \, \lambda\mu_{m'_{e}m_{g}} \, \mathcal{I}(\omega_{\text{ph}}+m_{g},\kappa) \notag \\
    &\quad \times \lambda\mu_{m_{g}m_{e}} \, \mathcal{I}(\omega_{e}+m_{e},\gamma) \, \lambda\sqrt{N}\mu_{m_{e}0_{g}}, \\   
       \bigg(\frac{\omega_{\text{ph}}}{2}\bigg)^3 \chi_N^{(5)}(\{\omega,\omega-\omega_{\text{ph}}\}^2,\omega) &= -\bm{V}_{0}\bm{G}_{e,0}\bm{v}_{0}\bm{G}_{\text{ph},1}\bm{v}_{0}^{\dagger}\bm{G}_{e,0}\bm{v}_{0}\bm{G}_{\text{ph},1}\bm{v}_{0}^{\dagger}\bm{G}_{e,0}\bm{V}_{0}^{\dagger} \notag \\
       &\quad - \bm{V}_{0}\bm{G}_{e,0}\bm{v}_{0}\bm{G}_{\text{ph},1}\bm{V}_{1}\bm{G}_{e,1}\bm{V}_{1}^{\dagger}\bm{G}_{\text{ph},1}\bm{v}_{0}^{\dagger}\bm{G}_{e,0}\bm{V}_{0}^{\dagger} \notag \\ 
     &= -\sum_{m_{e},m'_{e},m''_{e},m_{g},m'_{g}} \lambda\sqrt{N}\mu_{0_{g}m''_{e}} \, \mathcal{I}(\omega_{e}+m''_{e},\gamma) \, \lambda\mu_{m''_{e}m'_{g}} \, \mathcal{I}(\omega_{\text{ph}}+m'_{g},\kappa) \, \lambda\mu_{m'_{g}m_{e}} \notag \\
	&\quad \times \mathcal{I}(\omega_{e}+m'_{e},\gamma) \, \lambda\mu_{m'_{e}m_{g}} \, \mathcal{I}(\omega_{\text{ph}}+m_{g},\kappa) \, \lambda\mu_{m_{g}m_{e}} \, \mathcal{I}(\omega_{e}+m_{e},\gamma) \, \lambda\sqrt{N}\mu_{m_{e}0_{g}} \notag \\ 
	&\quad -\sum_{m_{e},m'_{e},m''_{e},m_{g}} \lambda\sqrt{N}\mu_{0_{g}m''_{e}} \, \mathcal{I}(\omega_{e}+m''_{e},\gamma) \, \lambda\mu_{m''_{e}m_{g}} \, \mathcal{I}(\omega_{\text{ph}}+m_{g},\kappa) \notag \\
	&\quad \times \bigg(\lambda\sqrt{N-1}\mu_{0_{g}m'_{e}} \, \mathcal{I}(\omega_{e}+m_{g}+m'_{e},\gamma) \, \lambda\sqrt{N-1}\mu_{m'_{e}0_{g}} \bigg) \notag \\
	&\quad \times \mathcal{I}(\omega_{\text{ph}}+m_{g},\kappa) \, \lambda\mu_{m_{g}m_{e}} \, \mathcal{I}(\omega_{e}+m_{e},\gamma) \, \lambda\sqrt{N}\mu_{m_{e}0_{g}}.
\end{align}
\end{subequations}

\noindent We return to equations for the second, fourth, and sixth order terms in the Dyson series of $D^R_N(\omega)$  and rewrite them in terms of the aforementioned nonlinear susceptibilities. 
\begin{center}
  \begin{tabular}{|c|c|c|}
\hline 
$D_{N}^{R,(2)}(\omega)$ & Dyson series terms & Using nonlinear susceptibilities\tabularnewline
\hline 
 & $\bm{G}_{\text{ph},0}\bigg(\bm{V}_0 \bm{G}_{e,0} \bm{V}^\dagger_0 \bigg)\bm{G}_{\text{ph},0}$ & $\bm{G}_{\text{ph},0}\big[ - \big(\frac{\omega_{\text{ph}}}{2}\big)\chi^{(1)}_N(\omega)\big]\bm{G}_{\text{ph},0}$ \tabularnewline
\hline 
\end{tabular} 
\end{center}

\begin{center}
  \begin{tabular}{|c|c|c|}
\hline 
$D_{N}^{R,(4)}(\omega)$ & Dyson series terms & Using nonlinear susceptibilities\tabularnewline
\hline 
 & $\bm{G}_{\text{ph},0}\cdot\bigg(\bm{V}_0\bm{G}_{e,0}\bm{V}_0^\dagger \bm{G}_{\text{ph},0}\bigg)^2$ & $\bm{G}_{\text{ph},0}\cdot\big(\big[ -\big(\frac{\omega_{\text{ph}}}{2}\big)\chi^{(1)}_N(\omega)\big] \bm{G}_{\text{ph},0}\big)^2$ \tabularnewline
\hline 
 & $\bm{G}_{\text{ph},0}\bigg(\bm{V}_0\bm{G}_{e,0}\bm{v}_0 \bm{G}_{\text{ph},1}\bm{v}_0^\dagger \bm{G}_{e,0}\bm{V}_0^\dagger\bigg)\bm{G}_{\text{ph},0}$ & $\bm{G}_{\text{ph},0}\big[ -\big(\frac{\omega_{\text{ph}}}{2}\big)^2\chi^{(3)}_N(\omega,\omega-\omega_{\text{ph}},\omega)\big]\bm{G}_{\text{ph},0}$ \tabularnewline
\hline 
\end{tabular} 
\end{center}

\begin{center}\scriptsize
    \begin{tabular}{|c|c|c|}
        \hline
        \( D^{R,(6)}_N(\omega) \) & Dyson series terms  &Using nonlinear susceptibilities \\ \hline
        & \(\bm{G}_{\text{ph},0} \left(\bm{V}_0 \bm{G}_{e,0} \bm{V}_0^\dagger \bm{G}_{\text{ph},0}\right)^3\) & $\bm{G}_{\text{ph},0}\big(\big[ -\big(\frac{\omega_{\text{ph}}}{2}\big)\chi^{(1)}_N(\omega)\big] 
\bm{G}_{\text{ph},0}\big)^3$\\ \hline
        
        & \(\bm{G}_{\text{ph,0}}(\bm{V}_{0} \bm{G}_{e,0} \bm{V}_{0}^{\dagger}) \bm{G}_{\text{ph,0}} \bigg(\bm{V}_{0} \bm{G}_{e,0} \bm{v}_{0} \bm{G}_{\text{ph},1} \bm{v}_{0}^{\dagger} \bm{G}_{e,0} \bm{V}_{0}^{\dagger}\bigg) \bm{G}_{\text{ph,0}}\) & $\bm{G}_{\text{ph,0}}\big[ -\bigg(\frac{\omega_{\text{ph}}}{2}\bigg)\chi^{(1)}_N(\omega)\big]\bm{G}_{\text{ph,0}}[ -\big(\frac{\omega_{\text{ph}}}{2}\big)^2\chi^{(3)}_N(\omega,\omega-\omega_{\text{ph}},\omega)\big]\bm{G}_{\text{ph,0}}$ \\ \hline
        
        & \(\bm{G}_{\text{ph,0}} \bigg(\bm{V}_{0} \bm{G}_{e,0} \bm{v}_{0} \bm{G}_{\text{ph},1} \bm{v}_{0}^{\dagger} \bm{G}_{e,0} \bm{V}_{0}^{\dagger}\bigg) \bm{G}_{\text{ph,0}} \bigg(\bm{V}_{0} \bm{G}_{e,0} \bm{V}_{0}^{\dagger}\bigg) \bm{G}_{\text{ph,0}}\) & $\bm{G}_{\text{ph,0}}[ -\big(\frac{\omega_{\text{ph}}}{2}\big)^2\chi^{(3)}_N(\omega,\omega-\omega_{\text{ph}},\omega)\big]\bm{G}_{\text{ph,0}}\big[ -\big(\frac{\omega_{\text{ph}}}{2}\big)\chi^{(1)}_N(\omega)\big]\bm{G}_{\text{ph,0}}$ \\ \hline
        
        & \(\bm{G}_{\text{ph,0}} \bigg(\bm{V}_{0} \bm{G}_{e,0} \bm{v}_{0} \bm{G}_{\text{ph},1} \bm{v}_{0}^{\dagger} \bm{G}_{e,0} \bm{v}_{0} \bm{G}_{\text{ph},1} \bm{v}_{0}^{\dagger} \bm{G}_{e,0} \bm{V}_{0}^{\dagger}\bigg) \bm{G}_{\text{ph,0}}\)  &\\ 
        
       & + &\normalsize{$\bm{G}_{\text{ph,0}}\big[ -\big(\frac{\omega_{\text{ph}}}{2}\big)^3\chi^{(5)}_N(\omega,\omega-\omega_{\text{ph}},\omega,\omega-\omega_{\text{ph}},\omega)\big]\bm{G}_{\text{ph,0}}$} \\
  
        & \(\bm{G}_{\text{ph,0}} \bigg(\bm{V}_{0} \bm{G}_{e,0} \bm{v}_{0} \bm{G}_{\text{ph,1}} \bm{V}_{1} \bm{G}_{e,1} \bm{V}_{1}^{\dagger} \bm{G}_{\text{ph,1}} \bm{v}_{0}^{\dagger} \bm{G}_{e,0} \bm{V}_{0}^{\dagger}\bigg) \bm{G}_{\text{ph,0}}\) &   \\ \hline
    \end{tabular} 
\end{center}\normalsize

\noindent Note that amongst the two terms contributing to $\big(\frac{\omega_{\text{ph}}}{2}\big)^3 \chi_N^{(5)}(\{\omega,\omega-\omega_{\text{ph}}\}^2,\omega)$, the first term is penalized by $1/N^2$ and the second term by $1/N$. All the nonlinear susceptibilities explicitly presented in Eq.~\ref{eq:susceptibilities} have been pictorially shown as ladder diagrams for a three-level system in Fig. \textcolor{blue}{3}, main text. \\

\noindent The above tables reveal that some of the terms at a given order can be factored into terms of lower-order molecular nonlinear susceptibilities. We call these DS-FDs/CUT-E diagrams as \emph{reducible} diagrams. The others are called \emph{irreducible} diagrams, and the corresponding molecular susceptibilities are the \emph{irreducible} nonlinear susceptibilities. We saw from the rules for the CUT-E diagram rules presented in Sec.~\ref{rules} that a complete diagram has to start from $\bm{H}_{\text{ph},0}$ and end at $\bm{H}_{\text{ph},0}$. This result leads to the fact that the diagrams that reach $\bm{H}_{\text{ph},0}$ at an intermediate step are always factorizable into diagrams of lower orders. This observation can be explicitly seen in the abovementioned examples and can be checked for higher-order terms in the Dyson series: the reducible terms contributing to $D_N^{R,(k)}(\omega)$ contain $\bm{G}_{ph,0}$ in between, while the irreducible ones only contain $\bm{G}_{ph,0}$ at the beginning and at the end. We will next show that the self-energy term in $D^R_N(\omega)$ is a sum over all the \emph{irreducible} nonlinear susceptibilities, up to some constants. \\

\subsection{The self-energy, $\Sigma_{e,0}$, is constituted by all the \emph{irreducible} nonlinear susceptibilities of an $N$ molecular ensemble} 
\noindent The self-energy term in $D_N^R(\omega)$, Eq.~\ref{DRw1} for an $N$ ensemble, according to the matrix continued fraction expansion, Eq.~\ref{eq:matrix_series}, is given as,
\begin{equation}\label{eq.self-energy_contd_frac}
\Sigma_{e,0}(\omega)=\bm{V}_{0}\frac{1}{\omega-\bm{H}_{e,0}-\bm{v}_{0}^{\dagger}\frac{1}{\omega-\bm{H}_{\text{ph,1}}+i\kappa/2-\bm{V}_{1}^{\dagger}\frac{1}{\frac{\ddots}{\omega-\bm{H}_{e,N-1}+i\gamma/2-\bm{v}_{N}^{\dagger}\frac{1}{\omega-\bm{H}_{\text{ph,}N}+i\kappa/2}\bm{v}_{N}}}\bm{V}_{1}}\bm{v}_{0}}\bm{V}_{0}^{\dagger}.
\end{equation}

\noindent On the other hand, we have the result that the sum over the \emph{irreducible} nonlinear susceptibilities computed diagrammatically using the CUT-E diagram and the rules, given as a nested summation, 
\begin{eqnarray}\label{eq.self-energy}
   \nonumber \sum_{l=0}^{\infty}\bigg(\frac{\omega_{\text{ph}}}{2}\bigg)^l\chi_N^{(2l+1)}(\{\omega,\omega-\omega_{\text{ph}}\}^l,\omega) &=& - \sum_{n_{e,0}=0}^{\infty}\sum_{n_{\text{ph},1}=0}^{\infty}\dots
\sum_{n_{\text{ph},N}=0}^{\infty}\bm{V}_{0}\bm{G}_{e,0}\bigg[\bm{v}_{0}\bm{G}_{\text{ph},1}\bigg[\bm{V}_{1}\bm{G}_{\text{e},1}\bigg[\dots[\bm{v}_{N}\cdot\\ 
&&
\bm{G}_{\text{ph},N}\bm{v}_{N}^\dagger\bm{G}_{e,N-1}]^{n_{\text{ph},N}}\dots\bigg]\bm{V}_{1}^\dagger\bm{G}_{\text{ph},1}\bigg]^{n_{\text{ph,}1}}\bm{v}_{0}^\dagger\bm{G}_{e,0}\bigg]^{n_{e,0}}\bm{V}_{0}^{\dagger}. \end{eqnarray}
\noindent This result can be explicitly checked by the reader by isolating the pathways that start and end at $\bm{H}_{\text{ph},0}$, and do not involve $\bm{H}_{\text{ph},0}$ in between, in other words, they must have $\bm{V}_0$ and $\bm{V}_0^\dagger$ only in the left and right, respectively (see Sec.~\ref{sec:FD-1}). We notice that series constitutes a nested matrix geometric progression. Repeatedly using Eq.~\ref{eq:matrix_series}, we arrive at the fact that Eq.~\ref{eq.self-energy} is equal to Eq.~\ref{eq.self-energy_contd_frac}. 

\section{$\mathcal{O}(N^{-k})$ expansion}\label{sec:1/N_expansion_supp}
\noindent Expanding $D_N^R(\omega)=\sum_{k=1}^\infty d_{n,k}(\omega)$ (see main text, $1/N$ \textbf{expansion} ) where $d_{n,k}(\omega) \propto 1/N^k$, we can explicitly find $d_{n,k}(\omega)$ for $k=0,1$ as follows.

\noindent \textit{a. $d_{N,0}(\omega)$:}\label{supp:collective} This term is proportional to $N^{0}$, so it must not have the participation of any $\bm{v}_k\propto 1/\sqrt{N}$,
\begin{align}\label{eq:coll}
d_{N,0}(\omega) &=\frac{1}{\omega-\omega_{\text{ph}}+i\kappa/2-\bm{V}_{0}\bm{G}_{e,0}\bm{V}_{0}^{\dagger}}\\
&=\sum_{n_{e,0}=0}^{\infty}\bm{G}_{\text{ph},0}\bigg(\bm{V}_{0}\bm{G}_{e,0}\bm{V}_{0}^{\dagger}\bm{G}_{\text{ph},0}\bigg)^{n_{e,0}},
\end{align}
\noindent used Eq.~\ref{eq:matrix_series} to go from the first to the second row, the latter indicates that $d_{N,0}$ is the sum over all possible paths in the CUT-E diagram restricted to the zeroth-order box. 

\noindent \textit{b. $d_{N,1}(\omega)$:} Truncating Eq.~\ref{eq.self-energy_contd_frac} at $\bm{H}_{\text{e,1}}$ gives an $\mathcal{O}(1/N)$ expression for $D_N^R(\omega)$, $\mathcal{D}_{{\mathcal{O}(1/N)}}(\omega)$
\begin{equation}
    \mathcal{D}_{{\mathcal{O}(1/N)}}(\omega)=\frac{1}{\omega-\omega_{\text{ph}}+i\kappa/2-\bm{V}_{0}\frac{1}{\omega-\bm{H}_{e,0}-\bm{v}_{0}^{\dagger}\frac{1}{\omega-\bm{H}_{\text{ph,1}}+i\kappa/2-\bm{V}_{1}^{\dagger}\frac{1}{\omega-\bm{H}_{e,1}+i\gamma/2}\bm{V}_{1}}\bm{v}_{0}}\bm{V}_{0}^{\dagger}}.
\end{equation}
\noindent Converting the continued fractions to summations and isolating the part of the summation \textbf{containing} $d_{N,1}(\omega)$, 
\begin{align*}
{\tilde{\mathcal{D}}}_{\mathcal{O}(1/N)}(\omega)  & =\sum_{n_{e,0}=0}^{\infty}\bm{G}_{\text{ph},0}\bigg(\bm{V}_{0}\bm{G}_{e,0}\bigg[\bm{I}+\underbrace{{\color{olive}\bm{v}_{0}\bm{G}_{\text{ph},1}}{\color{magenta}\bigg(\bm{I}-\bm{V}_{1}\bm{G}_{e,1}}{\color{magenta}\bm{V}_{1}^{\dagger}\bm{G}_{\text{ph},1}\bigg)^{-1}}{\color{olive}\bm{v}_{0}^{\dagger}\bm{G}_{e,0}}}_{\bm{X}}\bigg]\bm{V}_{0}^{\dagger}\bm{G}_{\text{ph},0}\bigg)^{n_{e,0}}.
\end{align*}
\noindent The collective term in the Green's function is given by the sum of the $X^0$ terms, which we can see yields  Eq.~\ref{eq:coll} for $d_{N,0}(\omega)$. To obtain the $+1/N$ order correction, $d_{N,1}(\omega)$, we focus on the terms
linear in $\bm{X}$.
\noindent Note that the terms $\bm{V}_{0}\bm{G}_{e,0}\bm{V}_{0}^{\dagger}\bm{G}_{\text{ph},0}$, $\bm{V}_{0}\bm{G}_{e,0}\bm{X}\bm{V}_{0}^{\dagger}\bm{G}_{\text{ph},0}$
, and $\bm{G}_{\text{ph},0}$ are complex numbers and hence commute.
Looking at a particular $n_{e,0}$, the term linear in $\bm{V}_{0}\bm{G}_{e,0}\bm{X}\bm{V}_{0}^{\dagger}\bm{G}_{\text{ph},0}$
is $n_{e,0}\big(\bm{V}_{0}\bm{G}_{e,0}\bm{V}_{0}^{\dagger}\bm{G}_{\text{ph},0}\big)^{n_{e,0}-1}\big(\bm{V}_{0}\bm{G}_{e,0}\bm{X}\bm{V}_{0}^{\dagger}\bm{G}_{\text{ph},0}\big)$.
\noindent Thus, 

\begin{align*}
d_{N,1}(\omega) & =\bm{G}_{\text{ph},0}\sum_{n_{e,0}=0}^{\infty}n_{e,0}\big(\bm{V}_{0}\bm{G}_{e,0}\bm{V}_{0}^{\dagger}\bm{G}_{\text{ph},0}\big)^{n_{e,0}-1}\big(\bm{V}_{0}\bm{G}_{e,0}\bm{X}\bm{V}_{0}^{\dagger}\bm{G}_{\text{ph},0}\big)\\
 & =\bm{G}_{\text{ph},0}\sum_{n_{e,0}=0}^{\infty}\frac{d}{d(\bm{V}_{0}\bm{G}_{e,0}\bm{V}_{0}^{\dagger}\bm{G}_{\text{ph},0})}\big(\bm{V}_{0}\bm{G}_{e,0}\bm{V}_{0}^{\dagger}\bm{G}_{\text{ph},0}\big)^{n_{e,0}}\big(\bm{V}_{0}\bm{G}_{e,0}\bm{X}\bm{V}_{0}^{\dagger}\bm{G}_{\text{ph},0}\big)\\
 &= d_{n,0}(\omega)\cdot\bm{V}_{0}\bm{G}_{e,0}\bm{v}_{0}\big(\omega-   \bm{H}_{\text{ph},1}+i\kappa/2 -\bm{V}_{1}\bm{G}_{e,1}\bm{V}_{1}^{\dagger}\big)^{-1}\bm{v}_{0}^{\dagger}\bm{G}_{e,0}\bm{V}_{0}^{\dagger}\cdot d_{n,0}(\omega).
\end{align*}

\noindent \textit{c. $d_{N,2}(\omega)$:} Truncating Eq.~\ref{eq.self-energy_contd_frac} at $\bm{H}_{e,2}$ gives $D^R_N(\omega)$ up to $\mathcal{O}(1/N^2)$, 
\begin{equation}
    \mathcal{D}_{\mathcal{O}(1/N^2})(\omega)=\frac{1}{\omega-\omega_{\text{ph}}+i\kappa/2-\bm{V}_{0}\frac{1}{\omega-\bm{H}_{e,0}-\bm{v}_{0}^{\dagger}\frac{1}{\omega-\bm{H}_{\text{ph,1}}+i\kappa/2-\bm{V}_{1}^{\dagger}\frac{1}{\omega-\bm{H}_{e,1}+i\gamma/2-\bm{v}_{1}^{\dagger}\frac{1}{\omega-\bm{H}_{\text{ph,2}}+i\kappa/2-\bm{V}_{2}^{\dagger}\frac{1}{\omega-\bm{H}_{e,2}+i\gamma/2}\bm{V}_{2}}\bm{v}_{1}}\bm{V}_{1}}\bm{v}_{0}}}
\end{equation}
\noindent Converting this expression to a summation and isolating the components \textbf{containing} $d_{N,2}(\omega)$,
\small{\begin{align}\label{Eq:1/N2-1}
{\tilde{D}}_{\mathcal{O}(1/N^2)}(\omega)
 & =\sum_{n_{e,0}=0}^{\infty}\bm{G}_{\text{ph},0}\bigg(\bm{V}_{0}\bm{G}_{e,0}\bigg[\bm{I}+{\color{olive}\bm{v}_{0}\bm{G}_{\text{ph},1}}{\color{magenta}\sum_{n_{e,1}}\bigg(\bm{V}_{1}\bm{G}_{e,1}}\big(\bm{I}+{\color{blue}\bm{Y}}\big){\color{magenta}\bm{V}_{1}^{\dagger}\bm{G}_{\text{ph},1}\bigg)^{n_{e,1}}}{\color{olive}\bm{v}_{0}^{\dagger}\bm{G}_{e,0}}+\bm{X^{2}}\bigg]\bm{V}_{0}^{\dagger}\bm{G}_{\text{ph},0}\bigg)^{n_{e0}},
\end{align}}\normalsize

\noindent where 
${\color{blue}\bm{Y} =  \bm{v}_{1} \bm{G}_{\text{ph},2} {\color{red} \left( \bm{I} - \bm{V}_{2} \bm{G}_{e,2} \bm{V}_{2}^{\dagger} \bm{G}_{\text{ph},2} \right)^{-1} } \bm{v}_{1}^{\dagger} \bm{G}_{e,1} }$. The approach to obtain $d_{N,2}(\omega)$ is to isolate the contributions \textbf{upto} linear order in ${\color{blue} \bm{Y}}$ from \[
{\color{magenta}\sum_{n_{e,1}}\bigg(\bm{V}_{1}\bm{G}_{e,1}}\big(\bm{I}+{\color{blue}\bm{Y}}\big){\color{magenta}\bm{V}_{1}^{\dagger}\bm{G}_{\text{ph},1}\bigg)^{n_{e,1}}}
\] and expand the $n_{e,0}$ exponent as a multinomial to isolate terms $\propto 1/N^2$. However, since $\bm{V}_{1} \bm{G}_{e,1} \bm{V}_{1}^{\dagger} \bm{G}_{\text{ph},1}$ and $\bm{V}_{1}\bm{G}_{e,1}\bm{Y}\bm{V}_{1}^{\dagger}\bm{G}_{\text{ph},1}$
\noindent are matrices in general; they do not commute, and hence, we
need to use the following result: Consider two
matrices $A$ and $B$. We have, 
\[
\big(A+B\big)^{n}=A^{n}+B^{n}+\bigg(A^{n-1}B+A^{n-2}BA+\dots+A^{n-k-1}BA^{k}+\dots+BA^{n-1}\bigg)+\dots.
\]

\noindent Thus the terms linear in $B$ can be written as, $\sum_{k}A^{n-k-1}BA^{k}$. Applying this to Eq.~\ref{Eq:1/N2-1}, we obtain
\begin{equation}
 {\tilde{D}}_{\mathcal{O}(1/N^2)}(\omega)\stackrel{\mathcal{O}(Y)}{=}\sum_{n_{e,0}=0}^{\infty}\bm{G}_{\text{ph},0}\bigg(\bm{V}_{0}\bm{G}_{e,0}\bigg[\bm{I}+\bm{X}+\bm{X}^{2}+\bm{K}\bigg]\bm{V}_{0}^{\dagger}\bm{G}_{\text{ph},0}\bigg)^{n_{e0}},
\end{equation}
\noindent where,
\begin{equation*}
 \bm{K=}\sum_{n_{e,1}}{\color{magenta}\sum_{k=0}^{n_{e,1}-1}{\color{olive}\bm{v}_{0}\bm{G}_{\text{ph},1}}[\bm{V}_{1}\bm{G}_{e,1}\bm{V}_{1}^{\dagger}\bm{G}_{\text{ph},1}]^{n_{e,1}-k-1}\cdot\bigg(\bm{V}_{1}\bm{G}_{e,1}{\color{blue}\bm{Y}}\bm{V}_{1}^{\dagger}\bm{G}_{\text{ph},1}\bigg)[\bm{V}_{1}\bm{G}_{e,1}\bm{V}_{1}^{\dagger}\bm{G}_{\text{ph},1}]^{k}}{\color{olive}\bm{v}_{0}^{\dagger}\bm{G}_{e,0}}
\end{equation*}
\noindent To obtain $d_{N,2}(\omega)\propto N^{-k}$, we need to isolate the terms quadratic in $\bm{X}$, $d_{N,2}^{(1)}$  and the terms linear in $\bm{Y}$, $d_{N,2}^{(2)}$. We have, 
\begin{align*}
d_{N,2}^{(1)}(\omega) & =\bm{G}_{\text{ph},0}\sum_{n_{e,0}=0}^{\infty}\begin{pmatrix}n_{e,0}\\
2
\end{pmatrix}\big(\bm{V}_{0}\bm{G}_{e,0}\bm{V}_{0}^{\dagger}\bm{G}_{\text{ph},0}\big)^{n_{e,0}-2}\big(\bm{V}_{0}\bm{G}_{e,0}\bm{X}\bm{V}_{0}^{\dagger}\bm{G}_{\text{ph},0}\big)^{2}+\\
 & \quad\bm{G}_{\text{ph},0}\sum_{n_{e,0}=0}^{\infty}n_{e,0}\big(\bm{V}_{0}\bm{G}_{e,0}\bm{V}_{0}^{\dagger}\bm{G}_{\text{ph},0}\big)^{n_{e,0}-1}\big(\bm{V}_{0}\bm{G}_{e,0}\bm{X}^{2}\bm{V}_{0}^{\dagger}\bm{G}_{\text{ph},0}\big)\\
 & =\bm{G}_{\text{ph},0}\bigg[\frac{d^{2}}{d\big(\bm{V}_{0}\bm{G}_{e,0}\bm{V}_{0}^{\dagger}\bm{G}_{\text{ph},0}\big)^{2}}\sum_{n_{e,0}=0}^{\infty}\big(\bm{V}_{0}\bm{G}_{e,0}\bm{V}_{0}^{\dagger}\bm{G}_{\text{ph},0}\big)^{n_{e,0}}\bigg]\big(\bm{V}_{0}\bm{G}_{e,0}\bm{X}\bm{V}_{0}^{\dagger}\bm{G}_{\text{ph},0}\big)^{2}+\\
 & \quad\bm{G}_{\text{ph},0}\bigg[\frac{d}{d(\bm{V}_{0}\bm{G}_{e,0}\bm{V}_{0}^{\dagger}\bm{G}_{\text{ph},0})}\sum_{n_{e,0}=0}^{\infty}\big(\bm{V}_{0}\bm{G}_{e,0}\bm{V}_{0}^{\dagger}\bm{G}_{\text{ph},0}\big)^{n_{e,0}}\bigg]\big(\bm{V}_{0}\bm{G}_{e,0}\bm{X}^{2}\bm{V}_{0}^{\dagger}\bm{G}_{\text{ph},0}\big)\\
 & =2\bm{G}_{\text{ph},0}\big(1-\bm{V}_{0}\bm{G}_{e,0}\bm{V}_{0}^{\dagger}\bm{G}_{\text{ph},0}\big)^{-3}\big(\bm{V}_{0}\bm{G}_{e,0}\bm{X}\bm{V}_{0}^{\dagger}\bm{G}_{\text{ph},0}\big)^{2}+\\
 & \quad\bm{G}_{\text{ph},0}\big(1-\bm{V}_{0}\bm{G}_{e,0}\bm{V}_{0}^{\dagger}\bm{G}_{\text{ph},0}\big)^{-2}\big(\bm{V}_{0}\bm{G}_{e,0}\bm{X}^{2}\bm{V}_{0}^{\dagger}\bm{G}_{\text{ph},0}\big).\\
\end{align*}

\noindent As for $d_{N,2}^{(2)}(\omega)$, the form of the $\bm{K}$ does not allow us to obtain a closed-form solution for $d_{n,2}(\omega)$. Similar challenges will be faced for $d_{n,k}(\omega)$ for $k>2$.

\bibliographystyle{apsrev4-1-custom}
\bibliography{linear_response}